\documentclass[12pt, epsfig]{article}                            
\usepackage{epsfig} 
\usepackage{amssymb}
\usepackage{amsfonts}
\usepackage{amsmath}
\usepackage{mathrsfs}
\usepackage{nicefrac}
\usepackage{color} 

\DeclareMathOperator{\de}{det}

\newcommand{\Tr}{\mathrm{Tr}}
\newcommand{\tr}{\mathrm{tr}}
\newcommand{\De}{\partial}

\newcommand{\exl}{\left<}
\newcommand{\exr}{\right>}

\newcommand{\Pf}{\mathrm{Pf}}
\newcommand{\m}{\mathfrak{m}}
\newcommand{\W}{\mathcal{W}}
\newcommand{\V}{\mathrm{v}}

\begin{document}

\begin{titlepage}
{\hfill     IFUP-TH/2011-08}
\bigskip
%\bigskip

\begin{center}
{\huge  {\bf
Confining vacua in SQCD, the Konishi anomaly and the Dijkgraaf-Vafa superpotential
 } }
\end{center}

\bigskip
\begin{center}
{\large  L. Di Pietro$^{1,2}$, S. Giacomelli$^{3,4}$   \vskip 0.10cm
 }
\end{center}

\begin{center}
{\it   \footnotesize
%Dipartimento di Fisica ``E. Fermi" -- Universit\`a di Pisa $^{(2)}$, \\
%   Largo Bruno Pontecorvo, 3, Ed. C, 56127 Pisa,  Italy \\
SISSA/ISAS -  Via Bonomea 265, I - 34136, Trieste, Italy $^{(1)}$, \\
Istituto Nazionale di Fisica Nucleare -- Sezione di Trieste $^{(2)}$, \\
    Via Valerio 2, I - 34127, Trieste, Italy \\
Scuola Normale Superiore - Pisa ,
 Piazza dei Cavalieri 7, Pisa, Italy $^{(3)}$\\
Istituto Nazionale di Fisica Nucleare -- Sezione di Pisa $^{(4)}$, \\
     Large Bruno Pontecorvo, 3, Ed. C, 56127 Pisa,  Italy
   }

\end {center}

\noindent
{\bf Abstract:}
In this paper we revisit the analysis of vacua in $\mathcal{N}=2$ SQCD with generic bare quark masses, softly broken by a mass term for the chiral superfield $\Phi$ in the adjoint representation of the gauge group. These vacua are labelled by an integer $r$ (r vacua) and can be studied at the semiclassical level by means 
of the equations of motion (for large mass) or nonperturbatively by means of the Seiberg-Witten curve (for small mass). Making use of the Konishi anomaly and 
of the Dijkgraaf-Vafa superpotential we are able to interpolate between these two limits and better understand the properties 
of these vacua. In particular, we clarify the origin of the two to one map that relates semiclassical to quantum vacua.  
\vfill

\begin{flushright}
  August 2011
\end{flushright}
%\begin{flushleft}
%\footnotesize{dipietro@sissa.it\\
%si.giacomelli@sns.it} 
%\end{flushleft}
\end{titlepage}

\bigskip

\hfill{}

\section{Introduction}

Supersymmetric theories are an interesting laboratory for understanding the properties of nonAbelian gauge theories at the nonperturbative level; 
a better control on the strongly coupled regime of Yang-Mills theory is required in order to uncover the dynamics underlying the longstanding problems 
of confinement and chiral symmetry breaking in QCD. One of the most important results in this sense is the Seiberg-Witten solution\cite{SWI},
\cite{SWII}, which encodes the infrared dynamics of $\mathcal{N}=2$ gauge theories and exhibits a striking relation between monopoles, confinement and chiral 
symmetry breaking; whether all this is at work in the real world QCD is still an open problem. 

The analysis carried out in these papers 
(which deal with $SU(2)$ gauge theories) has been generalised to $SU(N)$ in \cite{AF,APSH,Klemm,KlemmI} and the main propeties of the theory have been 
carefully analized in \cite{Hanany,HananyI,APS,KM}. In these latter papers a much richer structure has emerged, in which monopoles of nonAbelian kind rather than ordinary 't Hooft-Polyakov monopoles \cite{TP} play a key role \cite{ABEKY}. The precise analysis of these vacua (called r vacua) is hard to carry out  just by means 
of the Seiberg-Witten solution and seems to be deeply related to nonAbelian duality (see e.g. \cite{SD,KD}), which is not well understood yet. 
In particular the analysis of \cite {KM} is suited to study the properties of the r vacua for values of the bare quark masses which are very large (semiclassical regime), in which
they behave as Higgs vacua, or very small (nonperturbative regime), in which they are confining; it remains anyway very difficult to make precise predictions
about the ``intermediate range'', leaving some points unclear. 

The aim of this paper is to approach this problem by making use of the deep connection between these theories and their ``softly broken'' version, 
obtained adding a mass term $\frac{\mu}{2}\Tr\Phi^2$ for the adjoint chiral multiplet (actually the formalism we will use works for more general, 
not necessarily renormalizable superpotentials), lifting the moduli space of the $\mathcal{N}=2$ theory and leaving a finite number of vacua. This strategy has 
been adopted in \cite{GVY} to study the $SU(2)$ theory with one flavor: for $\mu>>\Lambda_{N_f}$ the natural approach is to decouple semiclassically 
the adjoint field. The low-energy theory is $\mathcal{N}=1$ SQCD with a quartic interaction between quarks. Taking into account the corrections due to 
nonperturbative gauge dynamics, the vacua can be found solving the stationarity equations for the quark superpotential. Holomorphicity in the $\mu$ 
parameter ensures that this description gives the same number of supersymmetric vacua and the same pattern of breaking of the global symmetry as 
 those  implied by the SW solution, more suitable for the $\mu<<\Lambda_{N_f}$ limit. This one-to-one correspondence can be established by means 
of an exact relation between chiral condensates derived from the Konishi anomaly \cite{K1,K2}. The analysis of the more interesting $SU(N)$ case along these lines requires the use of 
the Dijkgraaf-Vafa superpotential \cite{DVII,DV,DVI,DVIII}, which allows us to capture all the holomorphic data of the theory by means of a planar calculation 
in a matrix model, and of the generalized Konishi anomaly \cite{FDSW,FSWI,FSW}. These tools allow us also to identify fundamental phase invariants in SYM theories
\cite{FSWI,KO,FF,FFI} and to address nonperturbative investigations in $\mathcal{N}=1$ SQCD (see for instance \cite{AE,BH} and references therein).

In section 2 we shall review the construction proposed in \cite{GVY} and extend it 
to the other asymptotically free cases (2 and 3 flavors). In the subsequent  sections the general $SU(N)$ case will be analyzed, by  use of the generalized Konishi anomaly relations \cite{FDSW} and the Dijkgraaf-Vafa superpotential. In section 3 we shall derive the anomaly equations and the matrix model superpotential for the theory under consideration (the analysis
is similar to the one proposed in \cite{UN}) and see how it allows to recover the instanton induced superpotentials of \cite{ADS,SS,SI}. In section 4 
the previous results will be used  to derive some general features of the r vacua, in particular regarding the intermediate mass range and the ``transition'' from the Higgs to the pseudo-confining 
phase (as they call it in \cite{FSW}). As a byproduct we find  a clear interpretation of the two to one correspondence noted   in \cite{KM,BK} which 
associates both r and $N_f-r$ semiclassical vacua to the same quantum r vacuum.\\

\section{The anomaly technique for SU(2)}
As a warm-up, in this section we focus on an example with two colors and a renormalizable superpotential.  We consider a version of $\mathcal{N}=1$ SQCD with gauge group SU(2) and $N_f$ flavors, with in addition a chiral superfield $\Phi$ in the adjoint of the gauge group. The superpotential at the tree level is given by:
\begin{equation*}
\W_{tree}=m_i\tilde{Q}_iQ^i + \sqrt{2}h\tilde{Q}_i\Phi Q^i + \mu\Tr \Phi ^2.
\end{equation*}
$i=1,\ldots,N_{f}$ is a flavor index. Setting the Yukawa coupling $h=1$ and the adjoint mass $\mu=0$ we get $\mathcal{N}=2$ SQCD with $N_f$ flavors. For $N_f\leq 3$ this theory is asymptotically free and generates dynamically a scale which we denote by $\Lambda_{N_f}$. 
%\newline
%\indent Along the lines of \cite{GVY}, we will exploit the exact relations between chiral condensates derived from the Konishi anomaly to compare two different regimes of the theory:
%\begin{itemize}
%\item{for $\mu<<\Lambda_{N_f}$ and $h=1$ the low energy behavior is described by means of the exact solution of the $\mathcal{N}=2$ theory. The moduli space of this theory contains a Coulomb branch along which the gauge symmetry is broken to U(1). Adding the mass term $\mu$ for the field breaks the supersymmetry to $\mathcal{N}=1$ and lifts this flat direction. The vacua we are left with are only a finte number of points on the Coulomb branch, and possibly Higgs branches developing from them;}
%\item{for $\mu>>\Lambda_{N_f}$ after semiclassical decoupling of the adjoint field, the low energy theory is $\mathcal{N}=1$ SQCD with a quartic interaction between quarks. Taking into account the corrections due to nonperturbative gauge dynamics, the vacua can be found solving the stationarity equations for the quark superpotential.}
%\end{itemize}
The case with a single flavor was studied in \cite{GVY}, here we present the generalization to the other asymptotically free cases $N_f = 2, 3$. We show that even the description of the nontrivial flavor structure matches in the two regimes.

\subsection{Classical vacua and symmetries}
As explained in \cite{KM}, an analysis of the tree level superpotential reveals that for generic values of the quark and adjoint masses there are $N_f + 2$ supersymmetric vacua, and by a Witten index argument the number of vacua must be the same in the full quantum theory.  Flat directions develop when $\mu = 0$ or some of the quark masses coincide.  The 1-(complex) dimensional Coulomb branch of the $\mathcal{N}=2$ theory parametrized by $u=\langle \Tr \Phi ^2 \rangle$ is lifted by the soft breaking to $\mathcal{N} =1$ leaving only this discrete set of points. \newline
\indent The U(1)$_R$ and U(1)$_J \subset$ SU(2)$_R$ simmetries, together with holomorphicity, give constraints on the dependence of chiral condensates on the parameters in the superpotential. The charges are given in the following table:
\begin{center}
\begin{tabular}{|ccc|ccc|ccc|ccc|ccc|ccc|}
\hline
&& && $\Phi$ &&& $Q$ &&& $\tilde{Q}$ &&& $\mu$ &&& $m$ & \\
\hline
& U(1)$_R$ &&& 2 &&& 0 &&& 0 &&& -2 &&& 2 & \\
\hline
& U(1)$_J$ &&& 0 &&& 1 &&& 1 &&& 2 &&& 0 & \\
\hline
\end{tabular}
\end{center}
The U(1)$_R$ is anomalous at the quantum level, thus acting nontrivially on the dynamical scale $\Lambda_{N_f}$. The residual $\mathbb{Z}_{4(4-N_f)}$ symmetry is broken spontaneously to $\mathbb{Z}_4$ by the adjoint VEV leaving a $\mathbb{Z}_{4-N_f}$ acting on the u plane. \newline
\indent Alternatively we can define a modified and nonanomalous U(1)$_{R'}$ symmetry that acts on the Yukawa parameter $h$:
\begin{center}
\begin{tabular}{|ccc|ccc|ccc|ccc|ccc|ccc|ccc|}
\hline
&& && $\Phi$ &&& $Q$ &&& $\tilde{Q}$ &&& $\mu$ &&& $m$ &&& $h$ & \\
\hline
& U(1)$_{R'}$ &&& 1 &&& $\frac{N_f -2}{N_f}$ &&& $\frac{N_f -2}{N_f}$ &&& 0 &&& $\frac{4}{N_f}$ &&& $\frac{4-N_f}{N_f}$ &\\
\hline
\end{tabular}
\end{center}
The combination of the parameters that is neutral under the nonanomalous U(1)'s and adimensional is given by:
\begin{equation}\label{sigma}
 {\sigma_{N_f}}^2 \equiv (\Pf\,\m)^{1-\frac{4}{N_f}}\,h^4\,\Lambda_{N_f}^{4-N_f}
\end{equation}
where the mass matrix has been recast in the form:
\begin{equation*}
\m \equiv  \left( \begin{array}{c|c}
                                  0 & m\\
                                  \hline
                                  -m&0
                                  \end{array}\right), \qquad m=diag(m_1,\ldots , m_{N_f})
\end{equation*}
\indent The limit $\sigma_{N_f}\to 0$ is interesting since it can be interpret in two ways: either as the limit $h \to 0$ in which we recover $\mathcal{N}=1$ SQCD with massive quarks, or as the limit $\m \to \infty$ with $h=1$ in which the quarks decouple and the theory approaches pure $\mathcal{N}=2$ SYM softly broken by the adjoint mass term. Consistently, in both these limits we are left with two discrete supersymmetric vacua, respectively in the Higgs phase and in the confining phase. 
Notice that the Higgs and confining phases are continuosly connected in these theory: it displays complementarity, as expected for a theory with scalar fields in the fundamental.\newline
\indent When the $\mu$ parameter is large with respect to the dynamically generated scale, it is legitimate to study the low-energy theory by integrating out the adjoint field. As a result we get a version of $\mathcal{N}=1$ SQCD modified by a quartic term for the quarks:
\begin{equation*}
\W_{dec} = \frac{h^2}{8\mu}\tr[V^2] - \frac{1}{2}\tr[\m V]
\end{equation*}
where $V$ is the gauge invariant quark bilinear, assembled in a $2N_f \times 2N_f$ antisymmetric matrix, and $\tr$ denotes the trace over flavor indices. At the classical level $V$ is subject to the constraint $\Pf V = 0$. In the massless limit $\m=0$ the O(2$N_f$) flavor symmetry acts on $V$ by conjugation. It is convenient to express the $V$ matrix in term of neutral parameters $\V_1,\dots,\V_{N_f}$ as follows:
\begin{equation}\label{parametrizationV}
\exl V \exr \equiv \sigma_{N_f} \dfrac{\mu}{h^2} \left(\begin{array}{c|c} 
              0 & \begin{array}{ccc}
                      m_1\V_1  &  &                                    \\
                                & \ddots &                                \\
                                &   & m_{N_f}\V_{N_f}               \\
                     \end{array} \\
               \hline
              \begin{array}{ccc}
                        -m_1\V_1  &  &                                \\
                                & \ddots &                                \\
                                &   &   -m_{N_f}\V_{N_f}           \\
                     \end{array} & 0 
             \end{array}\right).
\end{equation}

\subsection{Low-energy effective superpotentials}

The superpotential does not receive any correction in perturbation theory but new terms may appear as a consequence of nonperturbative gauge dynamics. The results for the exact effective superpotentials in SQCD are known. For the theory in consideration it is only necessary to add the quartic term that is reminiscent of the microscopic Yukawa coupling to the $\Phi$. Therefore the general form is:
\begin{equation}\label{s.p.eff}
\W_{eff} = \W_{dec} + \W_{n.p.}
\end{equation}
We list below the results for the different values of $N_f$:
\begin{equation*}
\begin{array}{cl}
N_f=1 & \W_{n.p.} = {\tilde{\Lambda}_1}^{\phantom{1}5}(\mathrm{Pf}V)^{-1} \\
N_f=2 & \W_{n.p.} = X(\mathrm{Pf}V - \tilde{\Lambda}_2^{\phantom{1}4}) \\
N_f=3 & \W_{n.p.} = - \tilde{\Lambda}_3^{\phantom{1}-3}\mathrm{Pf}V.
\end{array}
\end{equation*}
In the second row $X$ is a nondynamical Lagrange multiplier implementing the modified quantum constraint. We denote by $\tilde{\Lambda}$ the dynamical scale in the theory where $\Phi$ has been decoupled. 

\subsection{The Konishi anomaly}
 In order to find where in the Coulomb branch the dicrete vacua are located we have to determine the adjoint field condensate $u=\langle \Tr \Phi ^2 \rangle$. This amounts to exploiting the anomalous Ward identities associated to Konishi anomaly: 
\begin{equation*}
\exl  X \frac{\partial \W}{\partial X} + T(R_X)\frac{\Tr W^2}{8\pi^2} \exr = 0
\end{equation*}
where $X$ is a generic chiral superfield, $T(R_X)$ is the Dynkin index of the representation of the gauge group acting on $X$. Specifyng $X$ to be $Q$, $\tilde{Q}$ or $\Phi$ we end up with:
\begin{equation*}
\left\{\begin{array}{l}
           Q^i, \tilde{Q}^i : \quad \exl \sqrt{2}h\tilde{Q}_{i} \Phi Q^{i} + m_i \tilde{Q}_i Q^i +\frac{1}{16\pi^2} \Tr W^2 \exr =0 \quad i=1,\ldots , N_f\\
           \Phi:\quad \exl 2\mu\Tr\Phi^2 +\sqrt{2}h\tilde{Q}_{i} \Phi Q^{i}+\frac{1}{4\pi^2}\Tr W^2\exr =0 .      
          \end{array}\right.
\end{equation*}
From this equations we eliminate the gaugino condensate  $s\equiv -\frac{1}{16\pi^2}\exl\Tr W^2\exr$ obtaining $s = \exl \frac{1}{2 N_f}\tr\left[\frac{h^2}{2\mu}V^2-\m V\right] \exr $. The remaining equations relates directly the interesting parameter $u$ to the quark VEVs:
\begin{equation}\label{Konishi}
 2 \mu u = \exl \dfrac{4-N_f}{2N_f}\tr\left[\frac{h^2}{2\mu} V^2\right] - \dfrac{2}{N_f}\tr\left[ \m V \right] \exr.
\end{equation}

\subsection{$N_f=1$}
%NOTA BENE: rispetto alle variabili $M$ e $\kappa$ dell'articolo di Gorsky Vanshtein Yung abbiamo $\V_1 = \dfrac{{m_1}^{\frac{3}{2}}}{\mu{\Lambda_1}^{\frac{3}{2}}} M = \kappa$%
In this section we review the results already obtained in \cite{GVY} with one flavor. Matching the running couplings at the scale $\mu$ we get: $\tilde{\Lambda}_1^{\phantom{1}5} = \mu^2{\Lambda_1}^3$. The stationarity of the superpotential $\eqref{s.p.eff}$ with respect to $V$ gives the equation:
\begin{equation}\label{stationarityNf=1}
\sigma_1\V _1 - 2 + 2 \dfrac{1}{{\V _1}^2} = 0.   
\end{equation}
where $\sigma_1 = h^2{m_1}^{-3/2}\Lambda_1^{3/2}$ and $\V_1 = (h^2\mu\Lambda_1^3)^{-1}m_1^2V_{12}$ are neutral under the nonanomalous U(1)'s.
Once this equation is solved, the corresponding value of $u$ is given by equation $\eqref{Konishi}$:
 \begin{equation}\label{KonishiNf=1}
4 u =(m\Lambda _1 ^3)^{1/2}( - 3\sigma_1{\V _1}^2 + 8 \V _1 )
\end{equation}

\paragraph{$\sigma_1\to 0$ limit:} the equation $\eqref{stationarityNf=1}$ has two solutions for finite values of the VEVs $\V_1 = \pm 1+ \mathcal{O}(\sigma _1)$ and one that goes to infinity $\V _1 = \frac{2}{\sigma _1} +  \mathcal{O}(\sigma _1)$. Correspondingly $\eqref{KonishiNf=1}$ gives:
\begin{equation*}
\left\{\begin{array}{l}
           \V _1 = \pm 1+ \mathcal{O}(\sigma _1) \Rightarrow u = (m\Lambda _1 ^3)^{1/2} (\pm 2 + \mathcal{O}(\sigma _1) ) \\
           \V _1 = \frac{2}{\sigma _1} + \mathcal{O}(\sigma _1) \Rightarrow u = (m\Lambda _1 ^3)^{1/2} (\frac{1}{\sigma_1} + \mathcal{O}(\sigma_1)) \\
          \end{array}\right.
\end{equation*}
From the point of view of the softly broken $\mathcal{N}=2$ theory the two vacua for finite $u$ are the monopole/dyon vacua while the third one is the electric charge vacuum that correctly goes to infinity in this limit.

\paragraph{$m_1\to 0$ limit:} since $\sigma_1 = \mathcal{O}(m^{-\frac{3}{2}})$ and $\V _1 = \mathcal{O}(m^{\frac{1}{2}})$ the equations become:
\begin{equation*}
\left\{\begin{array}{l}
\sigma _1 {\V _1}^3 + 2 = 0  \\
u = 2 (m\Lambda _1 ^3)^{1/2}  \V _1
\end{array}\right.\Rightarrow u = - 2 (m\Lambda _1 ^3)^{1/2} \left(\frac{2}{\sigma _1}\right)^{\frac{1}{3}} e^{\frac{2k\pi i}{3}},\, k=0,\,1,\,2 
\end{equation*}
Notice that we recover the $\mathbb{Z}_3$ symmetry acting on the $u$ plane, again in agreement with softly broken $\mathcal{N}=2$ SQCD with one massless flavor.

\paragraph{Seiberg-Witten curve:} the equivalence with softly broken $\mathcal{N}=2$ SQCD can be established for generic values of the parameters. If we eliminate $\V_1$ from equations $\eqref{stationarityNf=1}$ and $\eqref{KonishiNf=1}$ we get a single equation for the $u$ variable:
\begin{equation*}
h^2u^3 - {m_1}^2 u^2 - \frac{9}{2} m_1 {\Lambda_1}^3 h^2 u + 4 {m_1}^3 {\Lambda_1}^3 + \frac{27}{16} h^4 {\Lambda_1}^6 = 0.
\end{equation*}
Setting $h=1$ these equation is equivalent to the vanishing of the discriminant of the Seiberg-Witten curve for one flavor. The solutions correspond to the vacua of the softly broken theory with $\mu\neq 0$. Therefore the vacua obtained by our analysis are in one to one correspondence to those obtained by the Seiberg-Witten curve.

\subsection{$N_f=2$}
In this case the matching gives ${\tilde{\Lambda}_2}^{\phantom{1}2}=\mu\Lambda_2$. The stationarity of the effective superpotential $\eqref{s.p.eff}$ gives the equations:
\begin{equation}\label{stationarityNf=2}
\left\{\begin{array}{l}
           \sigma_2(\V_1 + \alpha X \V_2) - 2 = 0          \\
           \sigma_2(\V_2 + \alpha^{-1} X \V_1) - 2 = 0  \\
           \V_1\V_2 =  1
          \end{array}\right.
\end{equation}
in which $\sigma_2 = h^2 (m_1\,m_2)^{-1/2}{\Lambda_2}^2$, we have set $\alpha=m_2/m_1$ and a factor of $h^2/\mu$ has been reabsorbed in the definition of $X$. The variables $\V_{1,2}$ were defined in $\eqref{parametrizationV}$. The additional equation derived from the anomaly is:
\begin{equation}\label{KonishiNf=2}
4 u + \Lambda_2 (m_1\, m_2)^{1/2} [ \sigma_2 (\alpha^{-1} {\V_1}^2 + \alpha {\V_2}^2) - 4 (\alpha^{-1} \V_1 + \alpha \V_2) ]= 0.
\end{equation}
As a consistency check one can verify that in the decoupling limit of one of the two flavor, that is $m_2\to \infty$, the equations for the theory with a single flavor are correctly recovered.

\paragraph{$\sigma_2\to 0$ limit:} eliminating $X$, $\eqref{stationarityNf=2}$ has the two solutions $(\V_1,\,\V_2)=\pm(\alpha,\,\alpha^{-1}) + \mathcal{O}(\sigma _2)$ and correspondingly we get $u = \Lambda_2 (m_1\, m_2)^{1/2} (\pm 2 + \mathcal{O}(\sigma_2))$. These two solutions are the monopole/dyon vacua. One can easily see that there are two additional solutions going to infinity in the space of VEVs like $\sigma_2^{-1}$: these are vacua of the softly broken theory in which an electrically charged degree of freedom condensates. As explained above $\sigma_2\to 0$ can be interpreted as the limit in which quarks becomes very massive, and therefore it is correct for these vacua to go to infinity in this limit.

\paragraph{Degenerate masses $m_1=m_2=m$:} in this case $\alpha = 1$. The equations give the following solutions for $u$: 
\begin{eqnarray*}
 u_1 = u_2 =\Lambda_2 m [\sigma_2/2 + {\sigma_2}^{-1}] = \frac{1}{2} {\Lambda_2}^2 + m^2 , \\
  u_{3,4} =  \Lambda_2 m [- \sigma_2/2 \pm 2]= - \frac{1}{2} {\Lambda_2} ^2 \pm 2 m\Lambda_2.
\end{eqnarray*} 
The two coincident values $u_1$ and $u_2$ correspond to two different solutions for $(\V_1, \V_2)$, setting $m_1\neq m_2$ would split the degeneracy on the Coulomb branch and separate the two points. \newline
\indent When $m \to 0$ the full O(4) flavor symmetry is restored and in this limit the four vacua are organized in two pairs of coincident points on the Coulomb branch. This is again in agreement with the interpretation in terms of the softly broken $\mathcal{N}=2$ theory. From this point of view in each vacuum some charged degree of freedom condensates. The magnetic charges fall in spinorial representations of the flavor symmetry group and therefore in representations of Spin(4) $\simeq$ SU(2) $\times$ SU(2). The two pairs of coincident points correspond to two pairs of magnetically charged degrees of freedom that in the $m \to 0$ limit organize in a doublet of one of the two SU(2)s.

\subsection{$N_f=3$}
In this case the matching of the running couplings gives the relation: ${\tilde{\Lambda}_3}^{\phantom{1}3} = \mu^2 \Lambda_3 $. In terms of the $(\V_1, \V_2, \V_3)$ variables of $\eqref{parametrizationV}$ the stationarity conditions of the superpotential is:
\begin{equation}\label{stationarityNf=3}
\left\{ \begin{array}{l}
\sigma_3 \V _1 - 2 + 2{\alpha _1}^{-1} \V _2 \V _3 = 0 \\
\sigma_3 \V _2 -2 + 2{\alpha _2}^{-1} \V _3 \V _1 = 0  \\
\sigma_3 \V _3 -2 + 2{\alpha _3}^{-1} \V _1 \V _2 = 0 
\end{array}\right. 
\end{equation}
where $\sigma _3= h^2 {\tilde{\Lambda} _3}^{\phantom{1}\frac{3}{2}}\mu^{-1}(m_1 m_2 m_3)^{-\frac{1}{6}}$ and we defined $\alpha _1 = {m_1}^{\frac{4}{3}}(m_2 m_3)^{-\frac{2}{3}}$ with cyclic definitions for $\alpha_2$ and $\alpha_3$ such that $\alpha_1\alpha_2\alpha_3=1$. The adjoint field condensate is given by the anomaly equation:
\begin{equation*}
12 u + \sqrt{m_1\,m_2\,m_3 \,\Lambda_3}[\sigma _3 (\alpha _1 {\V _1}^2 + \alpha _2 {\V _2}^2 + \alpha _3 {\V _3}^2) - 8 (\alpha _1 \V _1 + \alpha _2 \V _2 + \alpha _3 \V _3)] = 0
\end{equation*}
Again one can check that in the limit $m_3 \to \infty$ , by imposing a correct scaling of the variables, the $N_f = 2$ equations are recovered, with $\V_3$ formally playing the role of the Lagrange multiplier. 

\paragraph{$\sigma_3\to 0$ limit:} the system $\eqref{stationarityNf=3}$ in this limit has two solutions for finite values of the parameters $(\V _1 ,\, \V _2,\, \V_3) = \pm({\alpha _1}^{-1} ,\, {\alpha _2}^{-1} ,\, {\alpha _3}^{-1}) + \mathcal{O}(\sigma _3)$ which give $u = \pm2\sqrt{m_1\,m_2\,m_3\,\Lambda_3}$. As for $N_f=1,2$ these are the two vacua which correspond to the monopole/dyon vacua. Again, one can find three additional solutions with a runaway $\sigma_3^{-1}$ behavior: these are in correspondence with the three vacua in the semiclassical region associated to the condensation of some electrically charged degree of freedom.

\paragraph{Degenerate masses $m_1=m_2=m_3=m$:} in this case $\alpha_1=\alpha_2=\alpha_3=1$. Solving the equations gives the following values for $u$:
\begin{eqnarray*}
 u_1 = u_2 = u_3 = \sqrt{\Lambda_3 m^3} [\sigma _3^{-1}+ \frac{1}{6} \sigma _3 - \frac{1}{16} {\sigma _3}^3]\\
 u_{4,5} = \sqrt{\Lambda_3 m^3} [-\frac{3}{4}\sigma _3 -\frac{1}{32}{\sigma _3}^3 \pm \frac{1}{4}\left(\frac{{\sigma _3}^2}{4} + 4\right)^{\frac{3}{2}}].
\end{eqnarray*}
Like in the $N_f = 2$ case, the fact that we find coincident points on the Coulomb brach is related to the partial restoration of the flavor symmetry. The full flavor symmetry in this case is O(6) and Spin(6) $\simeq$ SU(4) is explicitily broken by $m$ to SU(3)$\times$U(1). From the point of view of softly broken $\mathcal{N}=2$ there are three vacua associated to the condensation of magnetically charged degrees of freedom: when the masses are switched to have the same value, the three degrees of freedom organize in a fundamental multiplet of SU(3) and the vacua flow in the same point on the $u$ plane. The additional two solutions are singlets of SU(3). \newline
\indent In the limit $m \to 0$ the full O(6) is restored and monopoles/dyons should fall in multiplets of SU(4). This limit is equivalent to $\sigma_3\to\infty$ and indeed we see that the SU(3) triplet solution coincides with one of the two singlets in  $u = -\frac{1}{16}{\sigma _3}^3( 1 + \mathcal{O}({\sigma _3}^{-2}))$ and they form a fundamental multiplet of SU(4), while the other singlets move to $u=0$. Notice that there is no discrete symmetry acting on the $u$ plane when $N_f=3$. The presence of the singlet vacuum indicates that confinement can be realized even without dynamical breaking of the flavor symmetry.

\section{The $SU(N)$ theory}

We now consider the $SU(N)$ case with the same tree level superpotential as before. Decoupling the adjoint field we can express the effective superpotential
in terms of the meson field $M$, obtaining the result
\begin{equation}\label{potM}
\W(M,\Lambda)=\frac{1}{\mu}\left(\tr M^2-(\tr M)^2\right) + \tr mM + \W_{NP}, 
\end{equation}
where $\W_{NP}$ represents the nonperturbative contribution and its form depends on the range of $N_f$ as before:
%\begin{equation}\label{potenziali}
%\begin{array}{ccc}
% N_f<N_c & \qquad & \W_{NP}=\left(\frac{\Lambda_{1}^{3N_c-N_f}}{\de M}\right)^{\frac{1}{N_c-N_f}}\\
%N_f=N_c & \qquad & \W_{NP}=X\left(\de M - B\tilde{B} - \Lambda^{2N_c}\right)\\
%N_f=N_c + 1 & \qquad & \W_{NP}=\frac{1}{\Lambda_{1}^{2N_f-3}}\left[\de M-B^{i}M_{i}^{j}\tilde{B}_{j}\right]\\
%N_f>N_c + 1 & \qquad & \W_{NP}= qM\tilde{q}\\
%\end{array}
%\end{equation}
\begin{equation}\label{potenziali}
\begin{array}{|c|c|}
\hline
N_f<N_c & \W_{NP}=\left(\frac{\Lambda_{1}^{3N_c-N_f}}{\de M}\right)^{\frac{1}{N_c-N_f}} \\
\hline
N_f=N_c & \W_{NP}=X\left(\de M - B\tilde{B} - \Lambda^{2N_c}\right) \\
\hline
N_f=N_c+1 & \W_{NP}=\frac{1}{\Lambda_{1}^{2N_f-3}}\left[\de M-B^{i}M_{i}^{j}\tilde{B}_{j}\right] \\
\hline
N_f>N_c+1 & \W_{NP}= qM\tilde{q} \\
\hline
\end{array}
\end{equation}
The only new ingredient here is given by the last row, in which the superpotential is written in terms of dual field variables, according to Seiberg duality \cite{SD}.

\subsection{Generalized anomaly equations}

The techniques used so far can be applied in this case as well (see  for instance \cite{KM} for a detailed calculation). However, we will find more convenient
to use the matrix model superpotential introduced by Dijkgraaf and Vafa. First of all, in order to locate the vacua in the moduli 
space of the $\mathcal{N}=2$ theory we have to determine all the correlators $U_{i}=\frac{1}{i}\langle\Tr\Phi^i\rangle$  and then use the formula given in \cite{FSW} which
relates the SW curve with these quantities:
\begin{equation}\label{curva}
P_{N}(x)=x^{N}e^{-\sum_{i}\frac{U_i}{x^i}}+\Lambda^{2N-N_f}\frac{(x+m)^{N_f}}{x^N}e^{\sum_{i}\frac{U_i}{x^i}}. 
\end{equation}
To extract such information we need to consider the generalised anomaly equations introduced in \cite{FDSW} (their validity has been proven perturbatively
in \cite{FDSW} and nonperturbatively in \cite{NP}). If we consider the following transformations
on the matter fields
$$\begin{aligned}
  \delta\Phi &= \frac{1}{z-\Phi},\\
 \delta\Phi &= \frac{W_{\alpha}W^{\alpha}}{z-\Phi},\\
\delta Q_i &= \frac{1}{z-\Phi}Q_i,
  \end{aligned}
$$
we obtain the Ward identities \cite{SA} (the sum over flavors $i$ is implied)
\begin{equation}\label{angen}
\begin{aligned}
&\left\langle  \Tr\frac{\mu\Phi}{z-\Phi}\right\rangle+\left\langle \sqrt{2}\tilde{Q}^{i}\frac{1}{z-\Phi}Q_{i}\right\rangle =2R(z)T(z),\\
&\left\langle  \Tr\frac{\mu\Phi W_{\alpha}W^{\alpha}}{z-\Phi}\right\rangle = -32\pi^2 R^{2}(z),\\
&\left\langle \tilde{Q}^{i}\frac{\sqrt{2}\Phi+m_i}{z-\Phi}Q_{i}\right\rangle = N_f R(z).
\end{aligned}
\end{equation}
In the previous formula we have used the generating functions of chiral ring correlators:
$$ T(z)=\left\langle\Tr\frac{1}{z-\Phi}\right\rangle,\; R(z)=\frac{-1}{32\pi^2}\left\langle\Tr\frac{W_{\alpha}W^{\alpha}}{z-\Phi}\right\rangle,\; M(z)=\left\langle\tilde{Q}^{i}\frac{1}{z-\Phi}Q_{i}\right\rangle.$$
Once we have determined the above generating functions, we can recover all the correlation functions of operators in the 
chiral ring expanding them about infinity. The $\frac{1}{z^2}$ term of the first equation in (\ref{angen}) and the $\frac{1}{z}$ term of the third
one give the Konishi anomaly, all the others represent various generalizations.

 There is now an important point to consider: the above formulas are valid for $U(N)$ and we have to modify them a little to study the $SU(N)$ theory, since the tracelessness of $\Phi$ is not preserved by
the above listed variations. It is sufficient to modify them by adding a term proportional to the identity such that the condition $\Tr\Phi=0$ is preserved:
$$\begin{aligned}
  \delta\Phi &= \frac{1}{z-\Phi}-\frac{T(z)}{N}I,\\
 \delta\Phi &= \frac{W_{\alpha}W^{\alpha}}{z-\Phi}+\frac{32\pi^2}{N}R(z)I,\\
\delta Q_i &= \frac{1}{z-\Phi}Q_i,
  \end{aligned}
$$
This modification does not affect the anomaly since the identity does not
couple to gluons and the only correction to equations (\ref{angen}) arises due to the presence of the superpotential (the same idea already appeared in \cite{KS},\cite{ST}). 
Making use of the ``modified variations'' we find the Ward identities for the $SU(N)$ theory:
\begin{equation}\label{SUN}
\begin{aligned}
&\left\langle  \Tr\frac{\mu\Phi}{z-\Phi}\right\rangle+\left\langle \sqrt{2}\tilde{Q}^{i}\frac{1}{z-\Phi}Q_{i}\right\rangle - \frac{\sqrt{2}T(z)}{N}\left\langle \tilde{Q}^{i}Q_{i}\right\rangle =2R(z)T(z),\\
&\left\langle  \Tr\frac{\mu\Phi W_{\alpha}W^{\alpha}}{z-\Phi}\right\rangle + \frac{32\sqrt{2}\pi^2}{N}R(z)\left\langle \tilde{Q}^{i}Q_{i}\right\rangle = -32\pi^2 R^{2}(z),\\
&\left\langle \tilde{Q}^{i}\frac{\sqrt{2}\Phi+m_i}{z-\Phi}Q_{i}\right\rangle = N_f R(z).
\end{aligned}
\end{equation}
We can see from the above relations that the anomaly equations for the $SU(N)$ theory with superpotential $\frac{1}{2}\mu\Tr\Phi^2$ are equivalent
to those of the $U(N)$ theory but with a different superpotential ($\frac{1}{2}\mu\Tr\Phi^2-a\mu\Tr\Phi$), where 
\begin{equation}  
a \equiv \frac{\sqrt{2}}{N\mu}\langle\tilde{Q}^{i}Q_{i}\rangle 
\end{equation}   
and $S$ denotes   the gluino condensate. 
Taking this modification into account we obtain from the second equation (see \cite{FDSW}) the relation
\begin{equation}\label{rr}
R(z)=\frac{1}{2}\left(\mu (z-a)-\sqrt{\mu^2(z-a)^2-4S\mu}\right).
\end{equation}
An important point is that  our generating functions are actually defined on a double cover of the z-plane, which we can describe using the matrix 
model curve (a sphere in our case) $$\Sigma : y^2=\mu^{2}(z-a)^2-4\mu S=\mu^2 [(z-a)^2-\tilde{z}],$$ which has a single branch cut. The $R(z)$ function actually assumes the above form on the first sheet of this Riemann surface (the one visible semiclassically) and for large z it behaves like $R(z)\simeq S/z$, tending to zero for $z\rightarrow\infty$, where the theory can be studied semiclassically \cite{FDSW,FSW}. In this limit the cut closes and we recover the invisibility of the second sheet at the classical leel. On the second sheet the sign of the square root changes and the asymptotic behaviour is $R(z)\simeq W'(z)$. We will exploit this property in section 5.

The $M(z)$ and $T(z)$ generating functions have $N_f$ poles located at the points $z_i\equiv-\frac{m_i}{\sqrt{2}}$ (some on the first and the others on the second sheet of the double cover of the z-plane, depending on the vacuum we are considering) and a pole at infinity. 
Using the above solution for $R(z)$, we can now determine $M(z)$ and $T(z)$ following the derivation in \cite{FSW} (see also eqs. 2.8-2.9 in \cite{UN}). Since our main interest in this paper is 
the theory with equal bare masses for the flavors, we will set from now on 
\begin{equation}  z_i=-\frac{m}{\sqrt{2}}   \equiv \eta\;,     \qquad\forall i=1,\dots N_f.  
\end{equation}
 Notice that we are considering the
equal mass limit of the theory with nondegenerate masses
\begin{equation}\label{RTM}
\begin{aligned}
M(z)=&\frac{N_f[R(z)+\frac{\mu}{2}(a-\eta)]+\frac{N_f-2r}{2}\sqrt{\mu^2(a-\eta)^2-4S\mu}}{\sqrt{2}z + m},\\
T(z)=&\frac{\frac{N_f-2r}{2}\sqrt{\mu^2(a-\eta)^2-4S\mu}}{(z+m/\sqrt{2})\sqrt{\mu^2(z-a)^2-4S\mu}} + \frac{N_f/2}{z+m/\sqrt{2}}+\\
&\frac{\mu(N-N_f/2)}{\sqrt{\mu^2(z-a)^2-4S\mu}},
\end{aligned}
\end{equation}
where $r$ is the number of poles located on the first sheet.
We can now explicitly determine the matrix model (Dijkgraaf-Vafa) superpotential \cite{DV},\cite{DVI},\cite{DVII} (see also \cite{DVIII}) which
encodes the holomorphic data of the theory as in \cite{FSW} (the analogous calculation for the $U(N)$ theory has been done in \cite{UN}). This will be the subject of the next section. 

\subsection{The Dijkgraaf-Vafa superpotential}

The aim of this section is to determine the DV superpotential in the r vacua of the $\mathcal{N}=2$ $SU(N)$ SQCD (for the basic properties of such vacua see e.g.
\cite{APS},\cite{KM}) softly-broken by a mass term for the adjoint chiral multiplet; such vacua can be described in the "matrix model language'' by putting
 r poles on the first sheet of the double cover of the z-plane and the other $N_f-r$ on the second one (at least for large values of $m$, 
where a semiclassical analysis is reliable). We can follow closely the calculation done in \cite{UN}. The effective superpotential assumes the following form (for quadratic tree level superpotential as in this
case):
\begin{equation}\label{eff}
\begin{aligned}
\W_{DV}=&-\frac{1}{2}N\Pi-\frac{1}{2}\sum_{i,\;r_i=0}\Pi^{0}-\frac{1}{2}\sum_{i,\;r_i=1}\Pi^{1}+\frac{1}{2}(2N-N_f)\W(\Lambda_0)+\\
&\frac{1}{2}\sum_i \W(\eta)-(2N-N_f)\pi iS+S\log\left(\frac{\sqrt{2}^{N_f}\Lambda^{2N-N_f}}{\Lambda_{0}^{2N-N_f}}\right). 
\end{aligned}
\end{equation}
By $\sum_i$ we indicate the sum over flavors and by $\sum_{i,\;r_i=0}$, $\sum_{i,\;r_i=1}$ we mean the sum of the contributions from poles on the second and on the first sheets 
respectively. In the previous formula we have used the notation
\begin{equation}\nonumber
\begin{aligned}
\Pi =& 2\int_{\sqrt{\tilde{z}}+a}^{\Lambda_0}\mu\sqrt{(z-a)^2-4S/\mu}dz=\mu(\Lambda_0-a)^2-2S-2S\log\frac{(\Lambda_0-a)^2 \mu}{S},\\
\Pi^{0} =& -\int_{q}^{\Lambda_0}\mu\sqrt{(z-a)^2-4S/\mu}dz= \frac{-\mu(\Lambda_{0}-a)^{2}}{2}+2S\log(\Lambda_0-a)+\\
& 2S\left[\frac{1}{2}+\frac{\mu(\eta-a)}{4S}\sqrt{(\eta-a)^{2}-\frac{4S}{\mu}}-\log\left(\frac{\eta-a}{2}+\frac{1}{2}\sqrt{(\eta-a)^{2}-\frac{4S}{\mu}}\right)\right],\\
\Pi^{1}=& -\int_{\tilde{q}}^{\Lambda_0}\mu\sqrt{(z-a)^2-4S/\mu}dz= \frac{-\mu(\Lambda_{0}-a)^{2}}{2}+2S\log(\Lambda_0-a)+\\
& 2S\left[\frac{1}{2}-\frac{\mu(\eta-a)}{4S}\sqrt{(\eta-a)^{2}-\frac{4S}{\mu}}-\log\left(\frac{\eta-a}{2}-\frac{1}{2}\sqrt{(\eta-a)^{2}-\frac{4S}{\mu}}\right)\right],\\
\end{aligned}
\end{equation}
where $\Lambda_0$ is a UV cutoff, $q$ and $\tilde{q}$ are the positions of the poles on the first and on the second sheet respectively. Substituting back in (\ref{eff}) we obtain for a r vacuum the result 
\begin{eqnarray}\label{pote}
\W_{DV}&&=S\left[N+\log\left(\frac{\mu^{N}\Lambda^{2N-N_f}}{S^N}\sqrt{2}^{N_f}\right)\right] - N\mu\frac{a^2}{2}\nonumber\\
&&-rS\left[\frac{1}{2}+\frac{\mu\xi}{4S}\sqrt{\xi^2-\frac{4S}{\mu}}-\frac{\mu\xi^2}{4S}-\log\left(\frac{\xi}{2}+\frac{1}{2}\sqrt{\xi^{2}-\frac{4S}{\mu}}\right)\right]\\
&&-(N_{f}-r)S\left[\frac{1}{2}-\frac{\mu\xi}{4S}\sqrt{\xi^2-\frac{4S}{\mu}}-\frac{\mu\xi^2}{4S}-\log\left(\frac{\xi}{2}-\frac{1}{2}\sqrt{\xi^{2}-\frac{4S}{\mu}}\right)\right].\nonumber
\end{eqnarray}
where we have introduced  
\begin{equation}
\xi  \equiv  a-\eta= a+m/\sqrt{2}\;. 
\end{equation}
In the previous formula the parameter $a$ has the meaning of a Lagrange multiplier, so the next step is to impose the condition 
\begin{equation}\label{trace}
\frac{\De\W_{DV}}{\De a}=\frac{\mu}{2}\left[(N_{f}-2r)\sqrt{(a-\eta)^{2}-4S/\mu}-(2N-N_f)a-N_{f}\eta\right]=0
\end{equation}
which enforces $\Tr\Phi=0$, and substitute back in (\ref{pote}). 
%Notice that our convention for the square root is
%\begin{equation}\label{conv}\sqrt{(a-\eta)^{2}-4S/\mu}=(\eta-a)\sqrt{1-\frac{4S}{\mu(a-\eta)^2}}.\end{equation} We have used it in deriving (\ref{trace}) from (\ref{pote}).  

The r vacua of our theory are associated to the critical points of the effective superpotential with respect to $S$. From its variation we obtain the equation
\begin{eqnarray}
\frac{\De\W_{DV}}{\De S} &=& r\log\left(\frac{\xi}{2}+\frac{1}{2}\sqrt{\xi^{2}-\frac{4S}{\mu}}\right) + (N_{f}-r)\log\left(\frac{\xi}{2}-\frac{1}{2}\sqrt{\xi^{2}-\frac{4S}{\mu}}\right)\nonumber\\
&& + \log\left(\frac{\mu^{N}\Lambda^{2N-N_f}}{S^N}\sqrt{2}^{N_f}\right)=0\nonumber,
\end{eqnarray}
or (for $r\leq\frac{N_f}{2}$)
\begin{equation}\label{gluino}
(N_{f}-2r)\log\left(\frac{a-\eta}{2}-\frac{1}{2}\sqrt{(a-\eta)^{2}-\frac{4S}{\mu}}\right) + \log\left(\frac{\mu^{N-r}\Lambda^{2N-N_f}}{S^{N-r}}\sqrt{2}^{N_f}\right)=0. 
\end{equation}
From equations (\ref{trace}) and (\ref{gluino}) we can determine $a$ and $S$ and then all the chiral correlators of the theory from (\ref{RTM}) (plugging (\ref{trace}) in
(\ref{RTM}) we can get rid of the square roots in the numerator and rewrite it in the form)
\begin{equation}\label{RTMI}
\begin{aligned}
M(z)=&\frac{\mu Na + N_f R(z)}{\sqrt{2}z + m},\\
T(z)=&\frac{\mu(N-N_f/2)}{\sqrt{\mu^2(z-a)^2-4S\mu}} + \frac{N_f/2}{z+m/\sqrt{2}}-\\
&\frac{\mu Na\left(1-\frac{N_f}{2N}\right)-\frac{N_f}{2\sqrt{2}}\mu m}{(z+m/\sqrt{2})\sqrt{\mu^2(z-a)^2-4S\mu}}.
\end{aligned}
\end{equation}

\subsection{Instanton-induced superpotentials and the Konishi anomaly}

The value of the meson and gluino condensates can also be obtained using the technique proposed for $SU(2)$: we extremize (\ref{potM}) and then use the Konishi anomaly equation to determine $S$. Combining the two equations we deduce the relation (no sum over $i$)
\begin{equation}\label{ciao}
S=-M_{i}\frac{\De\W_{NP}}{\De M_i},
\end{equation}
where $M_i$ is a diagonal element of the meson matrix. Using this equation we can write the solution for the meson condensate in the form
$$M_i = \frac{1}{2}\left[\frac{m\mu}{2}+\frac{\Tr M}{N}\pm\sqrt{\left(\frac{m\mu}{2}+\frac{\Tr M}{N}\right)^2-2S\mu}\right],$$
r solutions with the minus sign ($M_{i}^{-}$) and the others with plus ($M_{i}^{+}$) for a r vacuum; the important thing is that it is not necessary to know the precise form of $\W_{NP}$. Notice that taking the sum over $i$ of the above relation we obtain precisely equation (\ref{trace}) since $a=\frac{\sqrt{2}}{N\mu}\Tr M$. We can now use equation (\ref{gluino}) to derive a relation between $S$ and $M$:
\begin{eqnarray}
&&\log\left(\frac{S^N}{\mu^{N}\Lambda^{2N-N_f}\sqrt{2}^{N_f}}\right)= (N_{f}-r)\log\left(\frac{\xi}{2}-\frac{1}{2}\sqrt{\xi^{2}-\frac{4S}{\mu}}\right) + \nonumber\\
&& r\log\left(\frac{\xi}{2}+\frac{1}{2}\sqrt{\xi^{2}-\frac{4S}{\mu}}\right)=N_f\log\frac{S}{\mu} -(N_f -r)\log\left(\frac{\xi}{2}+\frac{1}{2}\sqrt{\xi^{2}-\frac{4S}{\mu}}\right)-\nonumber\\
&& r\log\left(\frac{\xi}{2}-\frac{1}{2}\sqrt{\xi^{2}-\frac{4S}{\mu}}\right)=N_f\log\frac{S}{\mu} -\log\left[ \left(\frac{\sqrt{2}}{\mu}\right)^{N_f}(M_{i}^{-})^{r}(M_{i}^{+})^{N_f -r}\right].\nonumber
\end{eqnarray}
In the last term we recognize the determinant of the meson matrix. We can thus rewrite the previous relation in the form
\begin{equation}\label{olo}
S^{N-N_f}=\frac{\mu^{N}\Lambda^{2N-N_f}}{\de M}=\frac{\Lambda_{1}^{3N-N_f}}{\de M}.
\end{equation}
Taking into account (\ref{ciao}) we can deduce from this equation that
\begin{itemize}
\item For $N_f < N$ $\W_{NP}$ is precisely the ADS superpotential given in (\ref{potenziali}).
\item For $N_f = N$ we have the constraint $\de M=\Lambda_{1}^{2N}$.
\item For $N_f = N+1$ $\W_{NP}$ assumes the form $\de M/\Lambda_{1}^{2N_f-3}$.
\item For $N_f > N+1$ we find the ``continuation'' of the ADS superpotential (the functional dependence on the fields is the same), which can be obtained from the superpotential given in (\ref{potenziali}) by integrating out the dual quarks (this is legal when the meson matrix has maximal rank); see \cite{KM} for a detailed discussion.
\end{itemize}
We see that we can easily recover the nonperturbative part of the superpotential using (\ref{pote}) without having to discuss the various ranges of $N_f$ separately (an analogous result holds for the $U(N)$ theory, as shown in \cite{OO}). The result obtained agrees with (\ref{potenziali}) once the massive fields have been integrated out and the baryons set to zero (we are discussing nonbaryonic vacua).
Such a relation between the DV superpotential and the instanton superpotentials (\ref{potenziali}) for $N_f\leq N$ has been noticed previously in \cite{YD},\cite{f=c}.
The derivation there is based on the relation (see e.g. \cite{UN},\cite{CV}) $$\W_{DV\; on-shell}=\frac{\mu}{2}\langle\Tr\Phi^2\rangle,$$ and on the explicit factorization of the Seiberg-Witten
curve using random matrices \cite{DJ}. Notice that the above relation is easily recovered from what we have done in the previous sections: the r.h.s. can be read off from the $1/z^3$ term of $T(z)$ in (\ref{RTMI}) and is equal to $S(N-\frac{N_f}{2})+\mu ma\frac{N}{2\sqrt{2}}$. The same expression can be obtained from (\ref{pote}), once (\ref{trace}) and (\ref{gluino}) are imposed.

\section{Chiral condensates in the r vacua and pseudo-confining phase}

In this section we use the machinery introduced so far to study the properties of r vacua. Our starting point is the system of equations (\ref{trace}) 
and (\ref{gluino}) and we will concentrate on the even $N_f$ case for simplicity (the other case is similar). The analysis performed so far is valid for 
large $m$, where semiclassical tools are reliable. 

The important point is that we can now let $m$ decrease and follow the r vacua in the nonperturbative region,
comparing with the analysis at small $m$ performed in \cite{KM} using the SW curve. This process is quite nontrivial and, as we will see, many interesting phenomena emerge. 

We will show that for particular values of $m$ the r vacua merge in a superconformal fixed point and that in many cases for $m$ small enough the r vacua cross the cut 
of the $\mathcal{N}=1$ curve, signaling that a perturbative analysis is no more adequate. This is the basic ingredient which will allow us to understand precisely the correspondence between semiclassical (large $m$) and quantum r vacua.  

\subsection{Coalescence of the r vacua}

Solving (\ref{trace}) and (\ref{gluino}) in the general case is a hard task. However, the equations simplify considerably in the case $r=\frac{N_f}{2}$, leading to the solution
\begin{equation}\label{rmax}
a=\frac{N_f}{2N-N_f}\frac{m}{\sqrt{2}}, \quad S=2^{\frac{N_f}{2N-N_f}}\omega^{k} \mu\Lambda^{2} \quad k=0,\dots,N-\frac{N_f}{2}-1.
\end{equation}
Here $\omega$ is the $(N-\frac{N_f}{2})$-th root of unity, giving the expected number of vacua. Once we have determined these quantities we can calculate all the chiral condensates and determine the position of the vacuum in the $\mathcal{N}=2$ moduli space using equations (\ref{RTM}) and (\ref{curva}). If we now tune appropriately the bare mass of the quarks the poles associated to the matter fields (located at $\eta$)
coalesce with one of the branch points of the $\mathcal{N}=1$ curve. Imposing this condition we find (for $k=0$) 
\begin{equation}\label{mcrit}
m=\pm2^{\frac{6N-2N_f}{4N-2N_f}}\frac{2N-N_f}{2N}\Lambda.
\end{equation}
One can notice that for these particular values of the masses something special happens: the solutions (\ref{rmax}) become solutions of (\ref{trace}) and (\ref{gluino})
for every r! Since the position in the $\mathcal{N}=2$ moduli space is uniquely determined by $a$ and $S$ we find that for every r branch one vacuum coalesces with the
one we have considered so far, giving a superconformal point (in the $\mu=0$ limit) characterized by a higher singularity of the SW curve. This is clear if
we look at the factorization equation \cite{CV}
\begin{equation}\label{fact}
\begin{aligned}
& P_N(z)^2-4\Lambda^{2N-N_f}(z+m)^{N_f}=H^2(z)F(z),\\
& \quad y^2=\mu^2[(z-a)^2-4S/\mu]=Q^2(z)F(z).
\end{aligned}
\end{equation}
Since $S$ is nonzero the $\mathcal{N}=1$ curve is not a square and is divided by $z+m$ if the above condition is satisfied. In a $r=\frac{N_f}{2}$ vacuum $H^2(z)$
contains at least a $(z+m)^{N_f}$ factor so the curve can be rewritten as $(z+m)^{N_f+1}G(z)$. If $N_f=2N-2$ this point coincides with the maximally singular
point of \cite{GS}. As we change the value of $m$ the r vacua separate again. 

\subsection{Transition from the pseudo-confining to the Higgs phase}

Following the discussion in \cite{UN} we will now study the process of passing poles through the cut of the matrix model curve. As pointed out in \cite{FSW}, this fact signals the transition from the Higgs to the pseudo-confining phase. To discuss this issue we set $r=0$, so that equation (\ref{gluino}) becomes
\begin{equation}\label{taglio}
\tilde{\eta}-\tilde{a}=-\omega_{N_f}^{k}\frac{\hat{S}^{\frac{N}{N_f}}}{\sqrt{2}}-\omega_{N_f}^{-k}\hat{S}^{1- \frac{N}{N_f}}\sqrt{2},\quad \hat{S}=\frac{S}{\mu\Lambda^2},\; \tilde{\eta}-\tilde{a}=\frac{(\eta-a)}{\Lambda}.
\end{equation}
On the other hand, from the reduced $\mathcal{N}=1$ curve we deduce that the two branch points linked by the cut are located at $$\tilde{z}=\tilde{a}\pm 2\sqrt{\hat{S}}.$$
These two equations represent the basic ingredient for our analysis (from now on we will omit the tilde and the hat). We consider moving the $N_f$ poles on top of each other 
(since we are considering the equal mass limit) from infinity on the second sheet (since we are considering the $r=0$ case) towards the origin along a line of constant phase $\theta$ on the complex z-plane $$\eta=Re^{i\theta}.$$
With a suitable phase redefinition (also of $\eta$) we can put (\ref{taglio}) in the form $$\eta-a=-\frac{S^{\frac{N}{N_f}}}{\sqrt{2}}-\sqrt{2}S^{1- \frac{N}{N_f}}.$$  

\subsubsection*{$N_f=N$}

Let us start from the simplest case $N_f=N$ in which the previous equation becomes $$a-\eta=\frac{S}{\sqrt{2}}+\sqrt{2},$$ 
and from (\ref{trace}) we obtain the relation $$a=-\frac{S}{\eta}.$$ In order to understand how the cut moves as we change
$\eta$, we can use the above equations to reexpress the position of the branch points as a function of $\eta$. We find
\begin{equation}\label{branch}
z=a\pm 2\sqrt{S}= \sqrt{2}\pm\sqrt{-4\sqrt{2}\eta}.
\end{equation}

From this equation we can see that, as long as we keep the phase of $\eta$ constant, the cut does not rotate and its lenght is proportional to 
$\sqrt{\vert \eta\vert}$. The important point to notice is that the branch cut and the line of constant phase (the dashed line in figure1) always intersect at distance $\sqrt{2}$
from the origin, so the poles can pass through the cut only when $\vert \eta\vert=\sqrt{2}$ (or equivalently when the length of the cut is $4\sqrt{2}$,
as one can easily see from (\ref{branch})). We thus find the following picture: starting from infinity on the second sheet, the poles cross the cut as we reach
$\vert m\vert=2\Lambda$ (notice that this is precisely the value found in (\ref{mcrit}), if we set $N_f=N$) and end up in the first sheet. When we reach the origin (massless case) the cut closes
up. If we start increasing $\vert \eta\vert$, as we pass the critical value seen before, the poles are kicked back to the second sheet (see figure1). We thus learn that
the vacuum is always in the pseudo-confining phase for large values of the mass! 

\begin{figure}
\centering{\includegraphics[width=\textwidth]{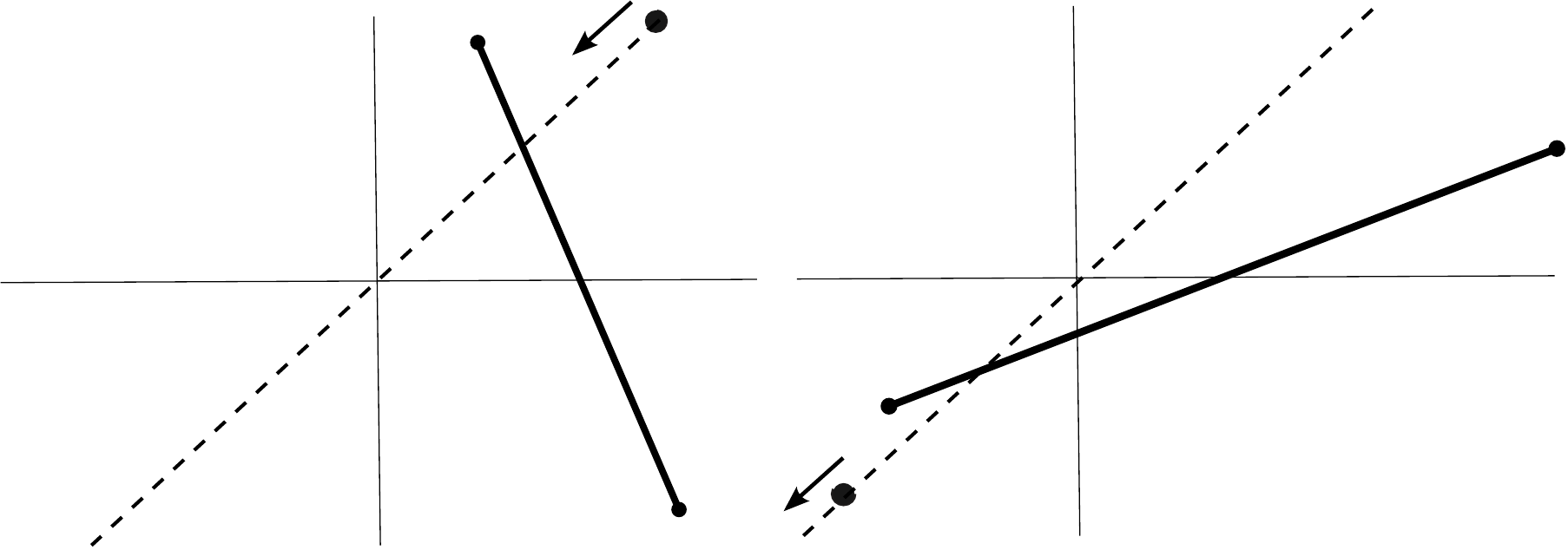}} 
\caption{\emph{(On the left) as the poles located at $\eta$ (represented by the dot on the dashed line) move towards the origin the cut (the thick line in the figure)
shrinks without rotating. (On the right) as the poles move from the origin on the first sheet to infinity they cross the cut (which opens up) and are sent to the second sheet.}}
\end{figure}

\subsubsection*{$N_f\neq N$}

Solving explicitly (\ref{taglio}) in this case is not simple, nonetheless we can deduce the basic features without a detailed calculation. Our expectation
is that for $N_f>N$ it is impossible to cross the cut only once and go to infinity on the first sheet, interpolating from the pseudo-confining to the Higgs
phase (see \cite{FSW} and \cite{UN}); we will now see that this is indeed the case. Let us recall our basic equations: (\ref{gluino})
\begin{equation}\label{asy}
\frac{S^q}{\sqrt{2}}=\frac{a-\eta}{2}\pm\frac{1}{2}\sqrt{(a-\eta)^2-4S},\quad q\equiv\frac{N}{N_f}, 
\end{equation}
where the + or - sign corresponds to a pole on the first or second sheet respectively (as we can read from \ref{gluino}) and (\ref{trace}) 
\begin{equation}
(2q-1)a+\eta=\mp \sqrt{(a-\eta)^2-4S} 
\end{equation}
(- for poles on the first sheet and + for poles on the second one). If we plug the second equation into the first one we obtain
$$\frac{S^q}{\sqrt{2}}=a(1-q)-\eta,$$ and comparing now with the second equation squared we find the relation
$$S^{1-q}=a\frac{q}{\sqrt{2}}.$$ Summing the last two relations we end up with 
$$\frac{S^q}{\sqrt{2}}+\sqrt{2}S^{1-q}=a-\eta.$$ We can deduce now that, for $N\leq N_f<2N$ (or $\frac{1}{2}<q\leq 1$), the asymptotic behaviour for 
large $\vert \eta\vert$ is $$\eta\simeq S^q,\quad a\simeq S^{1-q},$$ and this is incompatible with (\ref{asy}) taken with the + sign ,as we can see 
expanding the square root: the requirement that $R(z)$ vanishes on the first sheet for $z\rightarrow\infty$ (see the discussion after equation (\ref{rr})) leads to the following expansion for large mass 
\begin{equation}\label{conv}
\sqrt{\xi^2-4S}\simeq-\xi+\frac{2S}{\xi}+\dots
\end{equation}

We can conclude that, for $2N>N_f \geq N$, we cannot cross the cut once and go off to infinity on the first sheet. 
Clearly, this argument does not apply in the $N_f<N$ case: from (\ref{taglio}) we cannot conclude that S tends to infinity for large $\vert \eta\vert$ 
and the following asymptotic behaviour is allowed $$S\rightarrow 0, \quad \eta\simeq S^{1-q}.$$ This is compatible with (\ref{asy}), taken with the + sign.

\subsection{Generic r vacua: semiclassical analysis}

The equations for a generic r vacuum are not more difficult than those analized in the previous section so, the methods used there are applicable also
in this more general context. This section is devoted to the analysis of generic r vacua, resulting in a precise understanding of the relation between 
semiclassical and quantum vacua. The first thing one could ask is the following: from the analysis of the equations of motion (reliable in the large $m$
case) we can conclude that $r\leq\min{[N-1,N_f]}$. On the other hand, from what we have done so far, only the bound $r\leq N_f$ seems to be implied. Can
we recover the classical result from the matrix model framework? The answer is positive and we will now show it.

Let us consider a $r\geq N$ vacuum. Equations (\ref{trace}) and (\ref{gluino}) can be recast in the form (written in terms of the dimensionless variables
introduced in (\ref{taglio}))
\begin{equation}\label{key}
\begin{aligned}
&(N_{f}-2r)\sqrt{\xi^{2}-4S}=(2N-N_f)a+N_{f}\eta,\\
&\frac{\xi}{2}+\frac{1}{2}\sqrt{\xi^{2}-4S} = \frac{S^{1+\frac{N-r}{2r-N_f}}}{\sqrt{2}^{\frac{N_f}{2r-N_f}}}. 
\end{aligned} 
\end{equation}
If we now plug the first equation into the second one we obtain
\begin{equation}
\frac{S^{1+\frac{N-r}{2r-N_f}}}{\sqrt{2}^{\frac{3N_f-4r}{2r-N_f}}}=\xi\frac{2r-2N}{2r-N_f}+\eta\frac{2N}{N_f-2r}. 
\end{equation}
Squaring instead the second equation we find
\begin{equation}
\xi=\frac{S^{1+\frac{N-r}{2r-N_f}}}{\sqrt{2}^{\frac{N_f}{2r-N_f}}}+\sqrt{2}^{\frac{N_f}{2r-N_f}} S^{\frac{r-N}{2r-N_f}}. 
\end{equation}
Taking now the limit $\eta\rightarrow\infty$ (large $m$ or semiclassical limit) and recalling that $0\leq\frac{r-N}{2r-N_f}<\frac{1}{2}$ for $N_f<2N$ and 
$r\geq N$, we can deduce from these two equations that $$\xi\rightarrow\infty,\quad\xi\simeq S^{1+\frac{N-r}{2r-N_f}}.$$
As a consequence, we have that the ratio $S/\xi^2$ tends to zero and expanding the square root directly in (\ref{key}) we find, using (\ref{conv}) as before $$\xi\simeq S^{\frac{r-N}{2r-N_f}}.$$
Since we obtain two different asymptotic expansions for $\xi$, we can conclude that equations (\ref{key}) are inconsistent with the asymptotic behaviour of $R(z)$ given before, and we can discard r vacua
with r larger than (or equal to) N. We thus recover the semiclassical result.

We can understand in a similar way the semiclassical behaviour for general r. In the range $N_f-N_c<r<N$ (in the case $N_f>N$), depending on whether r is less or greater than
$\frac{N_f}{2}$, equation (\ref{gluino}) becomes respectively
$$\frac{\xi}{2}-\frac{1}{2}\sqrt{\xi^{2}-4S} = \frac{S^{\frac{N-r}{N_f-2r}}}{\sqrt{2}^{\frac{N_f}{N_f-2r}}},$$ or 
$$\frac{\xi}{2}+\frac{1}{2}\sqrt{\xi^{2}-4S} = \frac{S^{1+\frac{N-r}{2r-N_f}}}{\sqrt{2}^{\frac{N_f}{2r-N_f}}}.$$ We pass from the first to the second set
of vacua crossing the cut of the matrix model curve. In the first case, since $\frac{N-r}{N_f-2r}>\frac{1}{2}$
we have for large $\vert\eta\vert$ the following asymptotic behaviour from (\ref{conv}) $$S\rightarrow\infty,\quad\xi\simeq S^{\frac{N-r}{N_f-2r}}.$$
In the second case instead, from the fact that $\frac{N-r}{2r-N_f}>0$ we find $$S\rightarrow 0,\quad\xi\simeq S^{\frac{r-N}{2r-N_f}}.$$ 

A special role is played by vacua with $r=N_f-N$: in \cite{KM} it was argued that at the quantum level they are actually part of the baryonic root. The 
argument involves showing that for $m=0$ the SW curve becomes a perfect square, whereas for nonbaryonic vacua it is characterized by two single roots.
We will see in a moment that our formalism allows to recover such a result in a simple way. Equations (\ref{trace}) and (\ref{gluino}) become in this case
\begin{equation}
\begin{aligned}
&\left(\frac{\xi}{2}-\frac{1}{2}\sqrt{\xi^2-4S}\right)^{2N-N_f}=\frac{S^{2N-N_f}}{\sqrt{2}^{N_f}},\\
&\sqrt{\xi^2-4S}=\xi+\frac{2N}{2N-N_f}\eta.
\end{aligned}
\end{equation}
This system can be solved explicitly, leading to  the $2N-N_f$ solutions 
$$S=\sqrt{2}^{\frac{N_f}{2N-N_f}}\frac{N}{2N-N_f}\,\eta\,\omega^k,\quad\xi=-(\sqrt{2}^{\frac{N_f}{2N-N_f}}\omega^k+\frac{N}{2N-N_f}\eta),$$
where $\omega$ is the $2N-N_f$-th root of unity and $k=1,\dots,2N-N_f$ (note that in the special case $N=N_f$ we recover the result of the previous section).

The above solution tells us that $S$ vanishes in the massless case, recovering the result that the SW curve becomes a perfect square in that
limit, as we can see from the factorization equation (\ref{fact}). On the other hand, the vanishing of the gluino condensate signals a singularity in the 
description we are giving, as pointed out in \cite{UN}. This singularity is due to the presence of massless degees of freedom that we are missing; in 
this case they can be identified with the baryons, characteristic of the vacua in the baryonic root. 

The vacua with $r< N_f-N$ have the same asymptotic behaviour as those with $N_f-N<r<\frac{N_f}{2}$. The only difference is that crossing the cut we end 
up with a vacuum characterized by $r>N$, which does not exist semiclassically, as we have seen before. We conclude that it is not possible in this case to cross the cut once and 
then go off to infinity. A special case is given by the $r=0$ vacua discussed in the previous section.
It is anyway worth discussing them because the remaining baryonic vacua fall in this class. If we square equation (\ref{trace})
we find the relation
\begin{equation}\label{first}
S=\left(\frac{N-r}{N_f-2r}\xi+\frac{N}{N_f-2r}\eta\right)\left(\frac{N_f-N-r}{N_f-2r}\xi-\frac{N}{N_f-2r}\eta\right). 
\end{equation}
Plugging this into (\ref{gluino}) (and using (\ref{trace}) again) results in the equation
\begin{equation}
\begin{aligned}
&\sqrt{2}^{N_f}\left(\frac{N_f-N-r}{N_f-2r}\xi-\frac{N}{N_f-2r}\eta\right)^{N_f-2r}=\\
&\left(\frac{N-r}{N_f-2r}\xi+\frac{N}{N_f-2r}\eta\right)^{N-r}\left(\frac{N_f-N-r}{N_f-2r}\xi-\frac{N}{N_f-2r}\eta\right)^{N-r}.\\ 
\end{aligned}
\end{equation}
If $r>N_f-N$, then $N_f-2r<N-r$ and this equation has degree $2N-N_f$, giving the expected degeneracy of nonbaryonic vacua. If instead 
$r<N_f$ the previous equation becomes
$$\sqrt{2}^{N_f}\left(\frac{N_f-N-r}{N_f-2r}\xi-\frac{N}{N_f-2r}\eta\right)^{N_f-N-r}=\left(\frac{N-r}{N_f-2r}\xi+\frac{N}{N_f-2r}\eta\right)^{N-r}.$$
In this case the degree is $N-r=2N-N_f+(N_f-N-r)$. We thus find more solutions than the $2N-N_f$ associated to nonbaryonic roots;
these are precisely the missing baryonic vacua: in the limit $\eta\rightarrow0$ the above equation gives $N_f-N-r$ zero solutions. On the other
hand, from (\ref{first}) one can easily see that $S$ vanishes as well in this limit and the discussion made for the $r=N_f-N$
applies in this case too. The only difference is that the full $U(N_f)$ flavor symmetry is restored in this class of vacua (the vev
of the meson matrix vanishes in the massless limit).
 
In the theories with $N_f<N$ all the r vacua ($0\leq r\leq N_f$) fall in the class $N_f-N<r<N$ analysed above and also their asymptotic behaviour is the same,
so we do not need to discuss them further.

Let us summarize what we have found in this section:
\begin{itemize}
\item The r vacua exist for $0\leq r\leq\min{[N-1,N_f]}$.
\item For $N_f<N$ there are $2N-N_f$ vacua for every r in the above range.
\item For $r<N_f-N$ (so $N_f>N$) we have $2N-N_f$ nonbaryonic vacua and $N_f-N-r$ baryonic vacua characterized by a restoration 
of the flavor symmetry for every r.
\item For $r=N_f-N$ we have $2N-N_f$ baryonic vacua characterized by dynamical breaking of the flavor symmetry.
\item For $N_f-N<f<N$ we have found $2N-N_f$ nonbaryonic vacua for every r. 
\end{itemize}
In the nonbaryonic vacua the pattern of flavor symmetry breaking is $U(N_f)\rightarrow U(r)\times U(N_f-r)$. %If we deform the 
%quark masses taking them to be unequal, every r vacuum splits into ($\tiny\begin{array}{c}
%                                 N_f\\
%                                 r
%                                  \end{array}$) vacua, as shown in \cite{KM}. 
We thus recover precisely the vacuum counting performed there.

\subsection{Classical vs quantum r vacua}

Let us now move to the main result of this section. In \cite{KM,BK} it was noticed that there is a two-to-one correspondence, mapping both r and $N_f-r$ semiclassical vacua to r quantum vacua, which exist only for  $r \le  \tfrac{N_{f}}{2}$.  Making use of the matrix model technique we will be able to understand precisely the origin of this map.

Let us start with a r vacuum with $N_f-N<r\leq\frac{N_f}{2}$ and large $\vert\eta\vert$. If we now let the mass decrease our vacuum enters the nonperturbative region and can cross the cut of the matrix model curve. Depending on whether the vacuum crosses the cut or not, it is characterized
by r or $N_f-r$ poles on the second sheet respectively, and can be described by the (by now familiar) system of equations 
\begin{equation}\label{map}
\begin{aligned}
&\pm(2r-N_{f})\sqrt{\xi^{2}-4S}=(2N-N_f)\xi+2N\eta,\\
&\left(\frac{\xi}{2}\pm\frac{1}{2}\sqrt{\xi^{2}-4S}\right)^{N_f-2r} = \frac{S^{N-r}}{\sqrt{2}^{N_f}}. 
\end{aligned} 
\end{equation}
The sign is plus if it crosses the cut and minus in the other case. Our purpose is now to determine the locus on the mass plane on which our r vacuum crosses the cut. In order to do that we can add to the above system the equation 
\begin{equation}\label{cross}
\eta=a+2t\sqrt{S}\rightarrow\xi^2=4t^2S,\quad t\in[-1,1],
\end{equation}
and try to solve it. If we square the first equation in (\ref{map}), we find the relation 
$$4S=\xi^2-\left(\frac{2N-N_f}{N_f-2r}\xi+\frac{2N}{N_f-2r}\eta\right)^2.$$
Combining this with (\ref{cross}) we can rewrite $\eta$ in terms of $\xi$ and t:
\begin{equation}\label{eta}
\eta=\pm\frac{N_f-2r}{2Nt}\xi\left(\sqrt{t^2-1}\mp t\frac{2N-N_f}{N_f-2r}\right).
\end{equation}
Combining now the second equation in (\ref{map}) with the above relations we find
$$\left(\frac{\xi^2}{4t^2}\right)^{N-r}=\sqrt{2}^{N_f}\left(\frac{\xi}{2}\mp\frac{\sqrt{t^2-1}}{2t}\xi\right)^{N_f-2r}.$$ Notice that $N-r>N_f-2r$ since we are considering the case $r>N_f-N$, so we can simplify the equation and put it in the form (unless $\xi$ vanishes or equivalently $t=0$, but we will see that this is not a problem) $$\xi^{2N-N_f}=\sqrt{2}^{N_f}2^{2N-N_f}t^{2N-N_f}(t\mp\sqrt{t^2-1})^{N_f-2r}.$$
Using now (\ref{eta}) we can finally write the solution to our problem in the form
\begin{equation}\label{fund}
\eta=-\omega^k\frac{2N-N_f}{N}\sqrt{2}^{\frac{N_f}{2N-N_f}}(t\mp q\sqrt{t^2-1})(t\mp\sqrt{t^2-1})^{q},
\end{equation}
where $\omega$ is the $2N-N_f$-th root of unity, $k=1,\dots,2N-N_f$ and with $q$ we have indicated the ratio 
$$q\equiv\frac{N_f-2r}{2N-N_f}.$$ This is positive since $2r<N_f$ and $2N>N_f$ and is smaller than 1 because $r>N_f-N$.

In (\ref{fund}) we have found $4N-2N_f$ solutions: $2N-N_f$ is the number of r vacua and the sign ambiguity doubles that 
quantity. This is due to the fact that crossing the cut the r vacuum becomes a $N_f-r$ vacuum and (\ref{fund}) takes into 
account the contribution from both sets of vacua (notice that a change of sign in (\ref{fund}) can be undone sending 
$q\rightarrow-q$, or equivalently $r\rightarrow N_f-r$). Focusing now on a specific vacuum (we set $k=0$, the discussion 
is essentially unchanged in the other cases) we can study the locus of points we have just determined: the first property 
is $$\sqrt{2}^{\frac{N_f}{2N-N_f}}\frac{N_f-2r}{N}\leq\vert\eta\vert\leq\sqrt{2}^{\frac{N_f}{2N-N_f}}\frac{2N-N_f}{N},$$ the 
maximum is attained for $t=\pm1$ and the minimum for $t=0$. This tells us that we can cross the cut and therefore interpolate
 between $r$ and $N_f-r$ vacua only for values of $m$ low enough. Consequently, a semiclassical approach will always suggest 
us that we are dealing with vacua of different kind. Anyway, we are now in the position to compare the semiclassical and 
quantum behaviours: looking at (\ref{fund}) (let's say with the plus sign, the other case can be recovered simply by complex 
coniugation) we can see that, as $t$ goes from 1 to $-1$, the phase of $\eta$ changes by $(1+q)\pi$. The crucial point now 
is that $q<1$ in the range we are discussing, so the curve will not be closed! 

From this analysis we learn that, starting in the semiclassical regime with a r vacuum, depending on how we choose to change 
the value of $\eta$, we can reach the very strongly coupled region $m\simeq0$ either crossing the cut or not; in the second 
case we still have a vacuum with $r$ poles on the first sheet, but in the first one r and $N_f-r$ are interchanged and going 
back to infinity without crossing the cut again we can freely interpolate between the two sets of vacua. Notice that such a
process requires passing in the ``strongly coupled region" of the m-plane, where a fully quantum description is needed. The 
same considerations are also valid in a $r'=N_f-r$ vacuum, apart from the fact that we have to interchange the first case 
with the second one. The result is thus that the $2N-N_f$ solutions of (\ref{map}) 
\begin{figure}
\centering{\includegraphics[width=.3\textwidth]{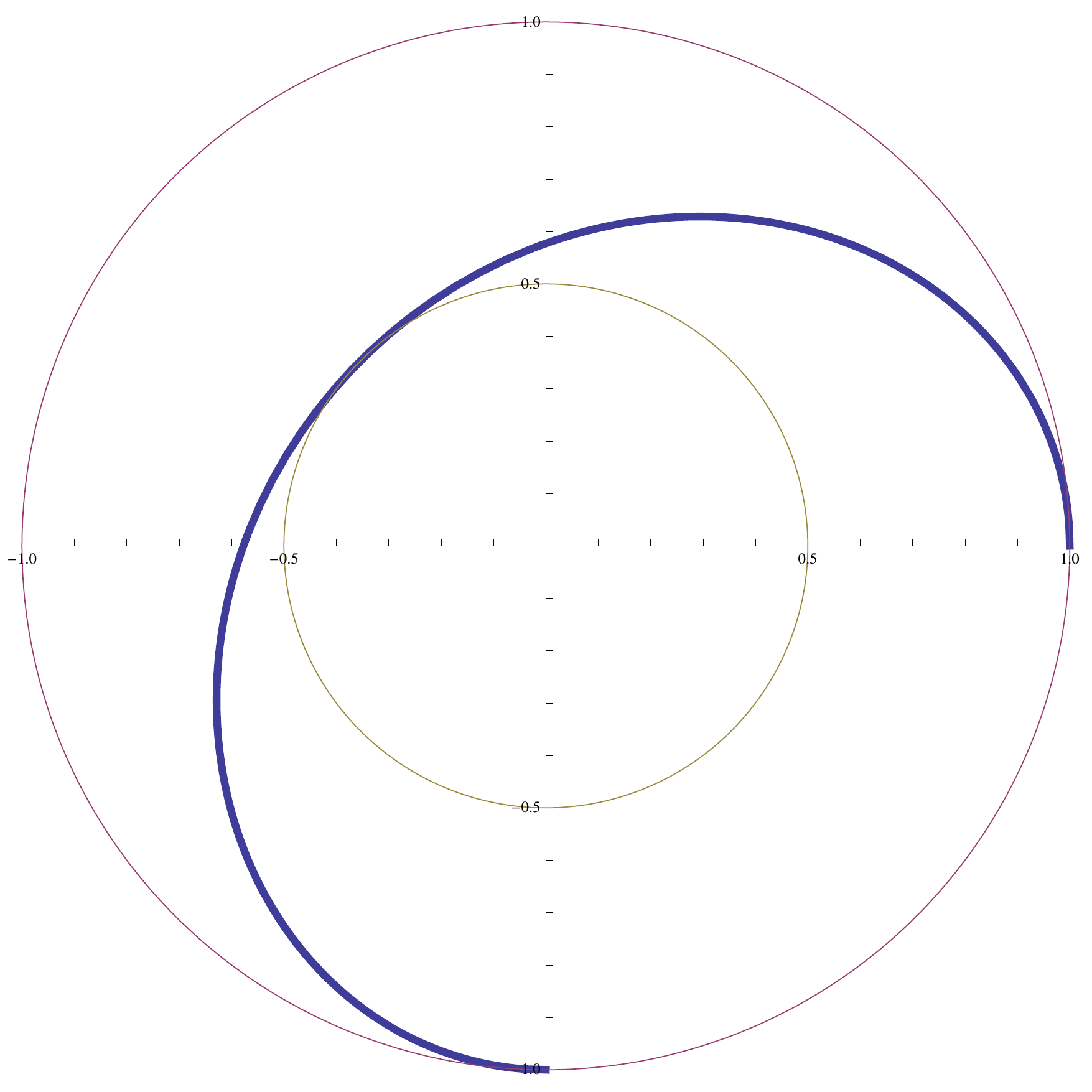}} 
\caption{\emph{We have plotted (\ref{fund}) on the complex plane for $q=\frac{1}{2}$ (the thick blue line). For $t=1$ the phase is 0, for $t=0$ it
is equal to $\frac{3}{4}\pi$ and for $t=-1$ it reaches $\frac{3}{2}\pi$. The line is open (as expected for $q<1$) and the phase of $\eta$ changes by 
$\frac{3}{2}\pi$. $\vert\eta\vert$ attains its minimum for $t=0$ and its maximum for $t=\pm1$.}}
\end{figure}
(considering together the + and - cases) 
can all be obtained starting from a vacuum with $r<\frac{N_f}{2}$ and will coincide with those found starting from a $r'$ 
vacuum. Since $\xi$ and $S$ determine uniquely the location of the vacuum in the $\mathcal{N}=2$ moduli space, we conclude that there 
is no actual distinction between  $r$ and $N_f-r$ vacua at the quantum level (of course when both exist). This nicely 
explains the results found in \cite{KM,BK} by matching the semi-classical and quantum vacua with the same flavor symmetry 
breaking pattern.

\section* {Acknowledgment} We are indebted to Ken Konishi for suggesting the problem and for stimulating discussions at various stages of this work.

%%%%%%%%%%%%%%%%%%%%%%%%%%%%%%%%

%%%%%%%%%%%%%%%%%%%%%%%%%%%%%%%%%  

\end{document}